\documentclass[aps,pre,twocolumn,showpacs]{revtex4}
\usepackage{epsfig}
\usepackage{times}
\usepackage{amsmath}
\begin{document}

\title{Fundamental and Real-World Challenges in Economics}

\author{Dirk Helbing$^{1,2,3}$ and Stefano Balietti$^1$}
\affiliation{$^1$ETH Zurich, CLU E1, Clausiusstr. 50, 8092 Zurich, Switzerland\\
$^2$Santa Fe Institute, 1399 Hyde Park Road, Santa Fe, NM 87501, USA\\
$^3$Collegium Budapest - Institute for Advanced Study, Szenth\'{a}roms\'{a}g u. 2, 1014 Budapest, Hungary}

\begin{abstract}
In the same way as the Hilbert Program was a response to the foundational crisis of mathematics \cite{Hilbert}, this article tries to formulate a research program for the socio-economic sciences. The aim of this contribution is to stimulate research in order to close serious knowledge gaps in mainstream economics that the recent financial and economic crisis has revealed. By identifying weak points of conventional approaches in economics, we identify the scientific problems which need to be addressed. We expect that solving these questions will bring scientists in a position to give better decision support and policy advice. We also indicate, what kinds of insights can be contributed by scientists from other research fields such as physics, biology, computer and social science. In order to make a quick progress and gain a systemic understanding of the whole interconnected socio-economic-environmental system, using the data, information and computer systems available today and in the near future, we suggest a multi-disciplinary collaboration as most promising research approach. 
\end{abstract}
%\pacs{02.50.Le, 87.10.Hk, 87.23.Ge}
\maketitle

\section{Introduction}\label{intro}

``How did economists get it so wrong?''. Facing the financial crisis, this question was brilliantly articulated by the Nobel prize winner of 2008, Paul Krugman, in the New York Times \cite{NYT}. A number of prominent economists even sees a failure of academic economics \cite{Dahlem}. Remarkably, the following declaration has been signed by more than 2000 scientists \cite{declaration}: ``Few economists saw our current crisis coming, but this predictive failure was the least of the field's problems. More important was the profession's blindness to the very possibility of catastrophic failures in a market economy ... the economics profession went astray because economists, as a group, mistook beauty, clad in impressive-looking mathematics, for truth ... economists fell back in love with the old, idealized vision of an economy in which rational individuals interact in perfect markets, this time gussied up with fancy equations ... Unfortunately, this romanticized and sanitized vision of the economy led most economists to ignore all the things that can go wrong. They turned a blind eye to the limitations of human rationality that often lead to bubbles and busts; to the problems of institutions that run amok; to the imperfections of markets---especially financial markets---that can cause the economy's operating system to undergo sudden, unpredictable crashes; and to the dangers created when regulators don't believe in regulation. ... When it comes to the all-too-human problem of recessions and depressions, economists need to abandon the neat but wrong solution of assuming that everyone is rational and markets work perfectly.''
\par
Apparently, it has not always been like this. DeLisle Worrell writes: ``Back in the sixties ... we were all too aware of the limitations of the discipline: it was static where the world was dynamic, it assumed competitive markets where few existed, it assumed rationality when we knew full well that economic agents were not rational ... economics had no way of dealing with changing tastes and technology ...  Econometrics was equally plagued with intractable problems: economic observations are never randomly drawn and seldom independent, the number of excluded variables is always unmanageably large, the degrees of freedom unacceptably small, the stability of significance tests seldom unequivocably established, the errors in measurement too large to yield meaningful results ...'' \cite{Lisle}.
\par
In the following, we will try to identify the scientific challenges that must be addressed to come up with better theories in the near future. This comprises practical challenges, i.e. the real-life problems that must be faced (see Sec. \ref{practical}), and fundamental challenges, i.e. the methodological advances that are required to solve these problems (see Sec. \ref{fundamental}). After this, we will discuss, which contribution can be made by related scientific disciplines such as econophysics and the social sciences. 
\par
The intention of this contribution is constructive. It tries to stimulate a fruitful scientific exchange, in order to find the best way out of the crisis. According to our perception, the economic challenges we are currently facing can {\it only} be mastered by large-scale, multi-disciplinary efforts and by innovative approaches \cite{FuturIcT}. We fully recognize the large variety of non-mainstream approaches that has been developed by ``heterodox economists''. However, the research traditions in economics seem to be so powerful that these are not paid much attention to. Besides, there is no agreement on which of the alternative modeling approaches would be the most promising ones, i.e. the heterogeneity of alternatives is one of the problems, which slows down their success. This situation clearly implies institutional challenges as well, but these go beyond the scope of this contribution and will therefore be addressed in the future.

\section{Real-World Challenges}\label{practical}

Since decades, if not since hundreds of years, the world is facing a number of recurrent socio-economic problems, which are obviously hard to solve. Before addressing related fundamental scientific challenges in economics, we will therefore point out practical challenges one needs to pay attention to. This basically requires to classify the multitude of problems into packages of interrelated problems. Probably, such classification attempts are subjective to a certain extent. At least, the list presented below differs from the one elaborated by Lomborg {\it et al.} \cite{Lomborg}, who identified the following top ten problems: air pollution, security/conflict, disease control, education, climate change, hunger/malnutrition, water sanitation, barriers to migration and trade, transnational terrorism and, finally, women and development. The following (non-ranked) list, in contrast, is more focused on socio-economic factors rather than resource and engineering issues, and it is more oriented at the roots of problems rather than their symptoms: 
\begin{enumerate}
\item {\it Demographic change} of the population structure (change of birth rate, migration, integration...)
\item {\it Financial and economic (in)stability} (government debts, taxation, and inflation/deflation; sustainability of social benefit systems; consumption and investment behavior...)
\item {\it Social, economic and political participation and inclusion} (of people of different gender, age, health, education, income, religion, culture, language, preferences; reduction of unemployment...)
\item {\it Balance of power} in a multi-polar world (between different countries and economic centers;
also between individual and collective rights, political and company power; avoidance of monopolies; formation of coalitions; protection of pluralism, individual freedoms, minorities...)
\item {\it Collective social behavior} and opinion dynamics (abrupt changes in consumer behavior; social contagion, extremism, hooliganism, changing values; breakdown of cooperation, trust, compliance, solidarity...)
\item {\it Security and peace} (organized crime, terrorism, social unrest, independence movements, conflict, war...)
\item {\it Institutional design} (intellectual property rights; over-regulation; corruption; balance between global and local, central and decentral control...)
\item {\it Sustainable use of resources and environment} (consumption habits, travel behavior, sustainable and efficient use of energy and other resources, participation in recycling efforts, environmental protection...) 
\item {\it Information management} (cyber risks, misuse of sensitive data, espionage, violation of privacy; data deluge, spam; education and inheritance of culture...) 
\item {\it Public health} (food safety; spreading of epidemics [flu, SARS, H1N1, HIV], obesity, smoking, or unhealthy diets...)
\end{enumerate}
Some of these challenges are interdependent.

\section{Fundamental Challenges}\label{fundamental}

In the following, we will try to identify the fundamental theoretical challenges that need to be addressed in order to understand the above practical problems and to draw conclusions regarding possible solutions. 

The most difficult part of scientific research is often not to find the right answer. The problem is to ask the right questions. In this context it can be a problem that people are trained to think in certain ways. It is not easy to leave these ways and see the problem from a new angle, thereby revealing a previously unnoticed solution. Three factors contribute to this:
\begin{enumerate}
\item We may overlook the relevant facts because we have not learned to see them, i.e. we do not pay attention to them. The issue is known from internalized norms, which prevent people from considering possible alternatives.
\item We know the stylized facts, but may not have the right tools at hand to interpret them.
It is often difficult to make sense of patterns detected in data. Turning data into knowledge is quite challenging. 
\item We know the stylized facts and can interpret them, but may not take them seriously enough, as we underestimate their implications. This may result from misjudgements or from herding effects, i.e. from a tendency to follow traditions and majority opinions. In fact, most of the issues discussed below have been pointed out before, but it seems that this did not have an effect on mainstream economics so far or on what decision-makers know about economics. This is probably because mainstream theory has become a norm \cite{Nelson}, and alternative approaches are sanctioned as norm-deviant behavior  \cite{vHayek-Individualism,vHayek-Science-Rev}.
\end{enumerate}
As we will try to explain, the following fundamental issues are not just a matter of approximations (which often lead to the right understanding, but wrong numbers). Rather they concern {\it fundamental} errors in the sense that certain conclusions following from them are seriously misleading. As the recent financial crisis has demonstrated, such errors can be very costly. However, it is not trivial to see what dramatic consequences factors such as dynamics, spatial interactions, randomness, non-linearity, network effects, differentiation and heterogeneity, irreversibility or irrationality can have. 

\subsection{Homo economicus}\label{economicus}

Despite of criticisms by several nobel prize winners such as Reinhard Selten (1994), Joseph Stiglitz and George Akerlof (2001), or Daniel Kahneman (2002), the paradigm of the {\it homo economicus}, i.e. of the ``perfect egoist'', is still the dominating approach in economics. It assumes that people would have quasi-infinite memory and processing capacities and would determine the best one among all possible alternative behaviors by strategic thinking (systematic utility optimization), and would implement it into practice without mistakes. The Nobel prize winner of 1976, Milton Friedman, supported the hypothesis of {\it homo economicus} by the following argument: ``irrational agents will lose money and will be driven out of the market by rational agents'' \cite{Friedman}. More recently, Robert E. Lucas Jr., the Nobel prize winner of 1995, used the  \textit{rationality} hypothesis to narrow down the class of empirically relevant equilibria \cite{Lucas}. 
\par
The rational agent hypothesis is very charming, as its implications are clear and it is possible to derive beautiful and powerful economic theorems and theories from it. The best way to illustrate {\it homo economicus} is maybe a company that is run by using optimization methods from operation research, applying supercomputers. Another example are professional chess players, who are trying to anticipate the possible future moves of their opponents. Obviously, in both examples, the future course of actions can not be fully predicted, even if there are no random effects and mistakes. 
\par
It is, therefore, no wonder that people have repeatedly expressed doubts regarding the realism of the rational agent approach \cite{Selten,Bounds}. Bertrand Russell, for example, claimed: ``Most people would rather die than think''. While this seems to be a rather extreme opinion, the following scientific arguments must be taken seriously:
\begin{enumerate}
\item Human cognitive capacities are bounded \cite{Simon,satisficing}. Already phone calls or conversations can reduce people's attention to events in the environment a lot. Also, the abilities to memorize facts and to perform complicated logical analyses are clearly limited.
\item In case of NP-hard optimization problems, even supercomputers are facing limits, i.e. optimization jobs cannot be performed in real-time anymore. Therefore, approximations or simplifications such as the application of heuristics may be necessary. In fact, psychologists have identified a number of heuristics, which people use when making decisions \cite{Gigerenzer}. 
\item People perform strategic thinking mainly in important new situations. In normal, everyday situation, however, they seem to pursue a satisficing rather than optimizing strategy \cite{satisficing}. Meeting a certain aspiration level rather than finding the optimal strategy can save time and energy spent on problem solving. In many situation, people even seem to perform routine choices \cite{Bounds}, for example, when evading other pedestrians in counterflows.
\item There is a long list of cognitive biases which question rational behavior \cite{List}.
For example, individuals are favorable of taking small risks (which are preceived as ``chances'', as the participation in lotteries shows), but they avoid large risks \cite{KahnemannTversky}. Furthermore, non-exponential temporal discounting may lead to paradoxical behaviors \cite{Didier} and requires one to rethink, how future expectations must be modeled.
\item Most individuals have a tendency towards other-regarding behavior and fairness \cite{Fairness,SmallScaleSocieties}. For example, the dictator game \cite{Dictator} and other experiments \cite{Ryan} show that people tend to share, even if there is no reason for this. Leaving a tip for the waiter in a restaurant that people visit only once is a typical example (particularly in countries where tipping is not common) \cite{Tipping}. Such behavior has often been interpreted as sign of social norms. While social norms can certainly change the payoff structure, it has been found that the overall payoffs resulting from them do not need to create a user or system optimum \cite{nonOptimal,Horne,withOpp}. This suggests that behavioral choices may be irrational in the sense of non-optimal. A typical example is the existence of unfavorable norms, which are supported by people although nobody likes them \cite{unfavorableNorms}. 
\item Certain optimization problems can have an infinite number of local optima or Nash equilibria, which makes it impossible to decide what is the best strategy \cite{MultipleEquilibria}.
\item Convergence towards the optimal solution may require such a huge amount of time that the folk theorem becomes useless. This can make it practically impossible to play the best response strategy \cite{Download}. 
\item The optimal strategy may be deterministically chaotic, i.e. sensitive to arbitrarily small details of the initial condition, which makes the dynamic solution unpredictable on the long run (``butterfly effect'') \cite{chaos,note}. This fundamental limit of predictability also implies a limit of control---two circumstances that are even more true for non-deterministic systems with a certain degree of randomness.
\end{enumerate}
In conclusion, although the rational agent paradigm (the paradigm of {\it homo economicus}) is theoretically powerful and appealing, there are a number of empirical and theoretical facts, which suggest deficiencies. In fact, most methods used in financial trading (such as technical analysis) are not well compatible with the rational agent approach.  Even if an optimal solution exists, it may be undecidable for practical reasons or for theoretical ones \cite{Goedel,TuringStoppingProblem}. This is also relevant for the following challenges, as {\it boundedly rational agents} may react inefficently and with delays, which questions the efficient market hypothesis, the equilibrium paradigm, and other fundamental concepts, calling for the consideration of spatial, network, and time-dependencies, heterogeneity and correlations etc. It will be shown that these points can have dramatic implications regarding the predictions of economic models.

\subsection{The efficient market hypothesis}\label{efficient}

The efficient market hypothesis (EMH) was first developed by Eugene Fama \cite{Fama} in his Ph.D. thesis and rapidly spread among leading economists, who used it as an argument to promote laissez-faire policies. The EMH states that current prices reflect all publicly available information and (in its stronger formulation) that prices instantly change to reflect new public information.
\par
The idea of self-regulating markets goes back to Adam Smith \cite{Wikipedia}, who believed that ``the free market, while appearing chaotic and unrestrained, is actually guided to produce the right amount and variety of goods by a so-called `invisible hand'.'' Furthermore, ``by pursuing his own interest, [the individual] frequently promotes that of the society more effectually than when he intends to promote it'' \cite{AdamSmith1}. For this reason, Adam Smith is often considered to be the father of free market economics. Curiously enough, however, he also wrote a book on ``The Theory of Moral Sentiments'' \cite{AdamSmith2}. ``His goal in writing the work was to explain the source of mankind's ability to form moral judgements, in spite of man's natural inclinations towards self-interest. Smith proposes a theory of sympathy, in which the act of observing others makes people aware of themselves and the morality of their own behavior... [and] seek the approval of the `impartial spectator' as a result of a natural desire to have outside observers sympathize with them'' \cite{Wikipedia}. Such a reputation-based concept would be considered today as indirect reciprocity \cite{indirectReci}. 
\par
Of course, there are criticisms of the efficient market hypothesis \cite{efficiency}, and the Nobel prize winner of 2001, Joseph Stiglitz, even believes that ``There is not invisible hand'' \cite{Joe}. The following list gives a number of empirical and theoretical arguments questioning the efficient market hypothesis:
\begin{enumerate}
\item Examples of market failures are well-known and can result, for example, in cases of monopolies or oligopolies, if there is not enough liquidity or if information symmetry is not given. 
\item While the concept of the ``invisible hand'' assumes something like an optimal self-organization \cite{NJP}, it is well-known that this requires certain conditions, such as symmetrical interactions. In general, however, self-organization does not necessarily imply system-optimal solutions. Stop-and-go traffic \cite{stopandgo} or crowd disasters \cite{crowddisasters} are two obvious examples for systems, in which individuals competitively try to reach individually optimal outcomes, but where the optimal solution is dynamically unstable. 
\item The limited processing capacity of boundedly rational individuals implies potential delays in their responses to sensorial inputs, which can cause such instabilities \cite{delay}. For example, a delayed adaptation in production systems may contribute to the occurrence of business cycles \cite{mitWitt}. The same applies to the labor market of specially skilled people, which cannot adjust on short time scales. Even without delayed reactions, however, the competitive optimization of individuals can lead to suboptimal individual results, as the ``tragedy of the commons'' in public goods dilemmas demonstrates \cite{Hardin,Fehr2}.
\item Bubbles and crashes, or more generally, extreme events in financial markets should not occur, if the efficient market hypothesis was correct (see next subsection). 
\item Collective social behavior such as ``herding effects'' as well as deviations of human behavior from what is expected from rational agents can lead to such bubbles and crashes \cite{Preis}, or can further increase their size through feedback effects \cite{Akerlof}. Cyclical feedbacks leading to oscillations are also known from the beer game \cite{beergame} or from business cycles \cite{mitWitt}.
\end{enumerate}

\subsection{Equilibrium paradigm}\label{equilibrium}

The efficient market paradigm implies the equilibrium paradigm. This becomes clear, if we split it up into its underlying hypotheses:
\begin{enumerate}
\item The market {\it can} be in equilibrium, i.e. there exists an equilibrium.
\item There is one and only one equilibrium.
\item The equilibrium is stable, i.e. any deviations from the equilibrium due to ``fluctuations'' or ``perturbations'' tend to disappear eventually.  
\item The relaxation to the equilibrium occurs at an infinite rate.
\end{enumerate}
Note that, in order to act like an ``invisible hand'', the stable equilibrium (Nash equilibrium) furthermore needs to be a system optimum, i.e. to maximize the average utility. This is true for coordination games, when interactions are well-mixed and exploration behavior as well as transaction costs can be neglected \cite{convention}. However, it is not fulfilled by so-called social dilemmas \cite{Hardin}.
\par
Let us discuss the evidence for the validity of the above hypotheses one by one: 
\begin{enumerate}
\item A market is a system of extremely many dynamically coupled variables. Theoretically, it is not obvious that such a system would have a stationary solution. For example, the system could behave periodic, quasi-periodic, chaotic, or turbulent \cite{Wolfgang,Puu,Lorenz,Frank,Day,Weidlich2,Rosser}. In all these cases, there would be no convergence to a stationary solution.
\item If a stationary solution exists, it is not clear that there are no further stationary solutions. If many variables are non-linearly coupled, the phenomenon of multi-stability can easily occur  \cite{multistability}. That is, the solution to which the system converges may not only depend on the model parameters, but also on the initial condition, history, or perturbation size. Such facts are known as path-dependencies or hysteresis effects and are usually visualized by so-called phase diagrams \cite{PLoSBiol}. 
\item In systems of non-linearly interacting variables, the existence of a stationary solution does not necessarily imply that it is stable, i.e. that the system will converge to this solution. For example, the stationary solution could be a focal point with orbiting solutions (as for the classical Lotka-Volterra equations \cite{Lotka}), or it could be unstable and give rise to a limit cycle \cite{Hopf} or a chaotic solution \cite{chaos}, for example (see also item 1). In fact, experimental results suggest that volatility clusters in financial markets may be a result of over-reactions to deviations from the fundamental value \cite{inSelten}. 
\item An infinite relaxation rate is rather unusual, as most decisions and related implemenations take time \cite{Gint,Marsili}. 
\end{enumerate}
The points listed in the beginning of this subsection are also questioned by empirical evidence.
In this connection, one may mention the existence of business cycles \cite{mitWitt} or unstable orders and deliveries observed in the experimental beer game \cite{beergame}. Moreover, bubbles and crashes have been found in financial market games \cite{Hommes}. Today, there seems to be more evidence against than for the equilibrium paradigm.
\par
In the past, however, most economists assumed that bubbles and crashes would not exist (and many of them still do). The following quotes are quite typical for this kind of thinking (from \cite{Vitting}):
In 2004, the Federal Reserve chairman of the U.S., Alan Greenspan, stated that the rise in house values was ``not enough in our judgment to raise major concerns''.
In July 2005 when asked about the possibility of a housing bubble and the potential
for this to lead to a recession in the future, the present U.S. Federal Reserve
chairman Ben Bernanke (then Chairman of the Council of Economic Advisors) said:
``It's a pretty unlikely possibility. We've never had a decline in
housing prices on a nationwide basis. So, what I think is more likely is that house
prices will slow, maybe stabilize, might slow consumption spending a bit. I don't
think it's going to drive the economy too far from it's full path though.''
As late as May 2007 Bernanke stated that the Federal Reserve
``do not expect significant spillovers from the subprime market to the rest of
the economy''.
\par
According to the classical interpretation, sudden changes in stock prices result from new information, e.g. from innovations (``technological shocks''). The dynamics in such systems has, for example, been described by the method of comparative statics (i.e. a series of snapshots). Here, the system is assumed to be in equilibrium in each moment, but the equilibrium changes adiabatically (i.e. without delay), as the system parameters change (e.g. through new facts). Such a treatment of system dynamics, however, has certain deficiencies:
\begin{enumerate}
\item The approach cannot explain changes in or of the system, such as phase transitions (``systemic shifts''), when the system is at a critical point (``tipping point'').
\item It does not allow one to understand innovations and other changes as results of an endogeneous system dynamics.
\item It cannot describe effects of delays or instabilities, such as overshooting, self-organization, emergence, systemic breakdowns or extreme events (see Sec. \ref{linear}).
\item It does not allow one to study effects of different time scales. For example, when there are fast autocatalytic (self-reinfocing) effects and slow inhibitory effects, this may lead to pattern formation phenomena in space and time \cite{Turing,Murray}. The formation of settlements, where people agglomerate in space, may serve as an example \cite{agglomeration,withTadek}. 
\item It ignores long-term correlations such as memory effects. 
\item It neglects frictional effects, which are often proportional to change (``speed'') and occur in most complex systems. Without friction, however, it is difficult  to understand entropy and other path-dependent effects, in particular irreversibility (i.e. the fact that the system may not be able to get back to the previous state) \cite{Mimkes}. For example, the unemployment rate has the property that it does not go back to the previous level in most countries after a business cycle \cite{unemployment}. 
\end{enumerate}

\subsection{Prevalence of linear models}\label{linear}

Comparative statics is, of course, not the only method used in economics to describe the dynamics of the system under consideration. As in physics and other fields, one may use a linear approximation around a stationary solution to study the response of the system to fluctuations or perturbations \cite{pattform}. Such a linear stability analysis allows one to study, whether the system will return to the stationary solution (which is the case for a stable [Nash] equilibrium) or not (which implies that the system will eventually be driven into a new state or regime). 
\par
In fact, the great majority of statistical analyses use {\it linear} models to fit empirical data (also when they do not involve time-dependencies). It is know, however, that linear models have special features, which are not representative for the rich variety of possible functional dependencies, dynamics, and outcomes. Therefore, the neglection of non-linearity has serious consequences:
\begin{enumerate}
\item As it was mentioned before, phenomena like multiple equilibria, chaos or turbulence cannot be understood by linear models. The same is true for self-organization phenomena or emergence.  Additionally, in non-linearly coupled systems, usually ``more is different'', i.e. the system may change its behavior fundamentally as it grows beyond a certain size. Furthermore, the system is often hard to predict and difficult to control (see Sec. \ref{control}). 
\item Linear modeling tends to overlook that a strong coupling of variables, which would show a normally distributed behavior in separation, often leads to fat tail distributions (such as ``power laws'') \cite{Sornette,Bunde}. This implies that extreme events are {\it much} more frequent than expected according to a Gaussian distribution. For example, when additive noise is replaced by multiplicative noise, a number of surprising phenomena may result, including noise-induced transitions \cite{Horsthemke} or directed random walks (``ratchet effects'') \cite{Reimann}.
\item Phenomena such as catastrophes \cite{Zeeman} or phase transition (``system shifts'') \cite{phasetrans}  cannot be well understood within a linear modeling framework. The same applies to the phenomenon of ``self-organized criticality'' \cite{SOC} (where the system drives itself to a critical state, typically with power-law characteristics) or cascading effects, which can result from network interactions (overcritically challenged network nodes or links) \cite{irgc,Battiston}. It should be added that the relevance of network effects resulting from the on-going globalization is often underestimated. For example, ``the stock market crash of 1987, began with a small drop in prices which triggered an avalanche of sell orders in computerized trading programs, causing a further price decline that triggered more automatic sales.'' \cite{Why}
\end{enumerate}
Therefore, while linear models have the advantage of being analytically solvable, they are often unrealistic. Studying non-linear behavior, in contrast, often requires numerical computational approaches. It is likely that most of today's unsolved economic puzzles cannot be well understood through linear models, no matter how complicated they may be (in terms of the number of variables and parameters) \cite{Wolfgang,Puu,Lorenz,Krugman,Frank,Day,Weidlich2,Sterman,Brock,Auyang,Salzano,DelliGatti,Lux,Rosser}. The following list mentions some areas, where the importance of non-linear interdependencies is most likely underestimated: 
\begin{itemize}
\item collective opinions, such as trends, fashions, or herding effects, 
\item the success of new (and old) technologies, products, etc., 
\item cultural or opinion shifts, e.g. regarding nuclear power, genetically manipulated food, etc., 
\item the ``fitness'' or competitiveness of a product, value, quality perceptions, etc., 
\item the respect for copyrights, 
\item social capital (trust, cooperation, compliance, solidarity, ...), 
\item booms and recessions, bubbles and crashes, 
\item bank panics,
\item community, cluster, or group formation.
\item relationships between different countries, including war (or trade war) and peace.
\end{itemize}

\subsection{Representative agent approach}\label{representative}

Another common simplification in economic modeling is the representative agent approach, which is known in physics as mean field approximation. Within this framework, time-dependencies and non-linear dependencies are often considered, but it is assumed that the interaction with other agents (e.g. of one company with all the other companies) can be treated as if this agent would interact with an average agent, the ``representative agent''. 
\par
Let us illustrate this with the example of the public goods dilemma. Here, everyone can decide whether to make an individual contribution to the public good or not. The sum of all contributions is multiplied by a synergy factor, reflecting the benefit of cooperation, and the resulting value is equally shared among all people. The prediction of the representative agent approach is that, due to the selfishness of agents, a ``tragedy of the commons'' would result \cite{Hardin}. According to this, everybody should free-ride, i.e. nobody should make a contribution to the public good and nobody would gain anything. However, if everybody would contribute, everybody could multiply his or her contribution by the synergy factor. This example is particularly relevant as society is facing a lot of public goods problems and would not work without cooperation. Everything from the creation of public infrastructures (streets, theaters, universities, libraries, schools, the World Wide Web, Wikipedia etc.) over the use of environmental resources (water, forests, air, etc.) or of social benefit systems (such as a public health insurance), maybe even the creation and maintainance of a commonly shared language and culture are public goods problems (although the last examples are often viewed as coordination problems). Even the {\it process} of creating public goods is a public good \cite{Olson}.
\par
While it is a well-known problem that people tend to make unfair contributions to public goods or try to get a bigger share of them, individuals cooperate much more than one would expect according to the representative agent approach. If they would not, society could simply not exist. In economics, one tries to solve the problem by introducing taxes (i.e. another incentive structure) or a ``shadow of the future'' (i.e. a strategic optimization over infinite time horizons in accordance with the rational agent approach) \cite{Axelrod,Future}. Both comes down to changing the payoff structure in a way that transforms the public good problem into another one that does not constitute a social dilemma \cite{withSergi}. However, there are other solutions of the problem. When the realm of the mean-field approximation underlying the representative agent approach is left and spatial or network interactions or the heterogeneity among agents are considered, a miracle occurs: Cooperation can survive or even thrive through correlations and co-evolutionary effects \cite{Nowak,Pacheco,withYu}. 
\par
A similar result is found for the public goods game with costly punishment. Here, the representative agent model predicts that individuals avoid to invest into punishment, so that punishment efforts eventually disappear (and, as a consequence, cooperation as well). However, this ``second-order free-rider problem'' is naturally resolved and cooperation can spread, if one discards the mean-field approximation and considers the fact that interactions take place in space or social networks \cite{PLoSBiol}. Societies can overcome the tragedy of the commons even without transforming the incentive structure through taxes. For example, social norms as well as group dynamical and reputation effects can do so \cite{Ostrom}. The representative agent approach implies just the opposite conclusion and cannot well explain the mechanisms on which society is built. 
\par
It is worth pointing out that the relevance of public goods dilemmas is probably underestimated in economics. Partially related to Adam Smith's belief in an ``invisible hand'', one often assumes underlying coordination games and that they would automatically create a harmony between an individually and system optimal state in the course of time \cite{convention}. However, running a stable financial system and economy is most likely a public goods problem. Considering unemployment, recessions always go along with a breakdown of solidarity and cooperation. Efficient production clearly requires mutual cooperation (as the counter-example of countries with many strikes illustrates). The failure of the interbank market when banks stop lending to each other, is a good example for the breakdown of both, trust and cooperation. We must be aware that there are many other systems that would not work anymore, if people would lose their trust:  electronic banking, e-mail and internet use, Facebook, eBusiness and eGovernance, for example. Money itself would not work without trust, as bank panics and hyperinflation scenarios show. Similarly, cheating customers by selling low-quality products or selling products at overrated prices, or by manipulating their choices by advertisements rather than informing them objectively and when they want, may create profits on the short run, but it affects the trust of customers (and their willingness to invest). The failure of the immunization campaign during the swine flu pandemics may serve as an example. Furthermore, people would probably spend more money, if the products of competing companies were better compatible with each other. Therefore, on the long run, more cooperation among companies and with the customers would pay off and create additional value.
\par
Besides providing a misleading picture of how cooperation comes about, there are a number of other deficiencies of the representative agent approach, which are listed below:
\begin{enumerate}
\item Correlations between variables are neglected, which is acceptable only for ``well-mixing'' systems. According to what is known from critical phenomena in physics, this approximation is valid only, when the interactions take place in high-dimensional spaces or if the system elements are well connected. (However, as the example of the public goods dilemma showed, this case does not necessarily have beneficial consequences. Well-mixed interactions could rather cause a breakdown of social or economic institutions, and it is conceivable that this played a role in the recent financial crisis.)
\item Percolation phenomena, describing how far an idea, innovation, technology, or (computer) virus spreads through a social or business network, are not well reproduced, as they depend on details of the network structure, not just on the average node degree \cite{Carlos}. 
\item The heterogeneity of agents is ignored. For this reason, factors underlying economic exchange, perturbations, or systemic robustness \cite{Heterogeneous} cannot be well described. Moreover, as socio-economic differentiation and specialization imply heterogeneity, they cannot be understood as emergent phenomena within a representative agent approach. Finally, it is not possible to grasp innovation without the consideration of variability. In fact, according to evolutionary theory, the innovation rate would be zero, if the variability was zero \cite{EigenFisher}. Furthermore, in order to explain innovation in modern societies, Schumpeter introduced the concept of the ``political entrepreneur'' \cite{Schumpeter}, an extra-ordinarily gifted person capable of creating disruptive change and innovation. Such an extraordinary individual can, by definition, not be modeled by a ``representative agent''.
\end{enumerate}
One of the most important drawbacks of the representative agent approach is that it cannot explain the fundamental fact of economic exchange, since it requires one to assume a heterogeneity in resources or production costs, or to consider a variation in the value of goods among individuals. Ken Arrow, Nobel prize winner in 1972, formulated this point as follows \cite{Arrow}: ``One of the things that microeconomics teaches you is that individuals are not alike. There is heterogeneity, and probably the most important heterogeneity here is heterogeneity of expectations. If we didn't have heterogeneity, there would be no trade.'' 
\par
We close this section by mentioning that economic approaches, which go beyond the representative agent approach, can be found in Refs. \cite{Gallegati,Kirman}.

\subsection{Lack of micro-macro link and ecological systems thinking}\label{micromacro}

Another deficiency of economic theory that needs to be mentioned is the lack of a link between micro- and macroeconomics. Neoclassical economics implicitly assumes that individuals make their decisions in isolation, using only the information received from static market signals. Within this oversimplified framework, macro-aggregates are just projections of some representative agent behavior, instead of the outcome of complex interactions with asymmetric information among a myriad of heterogeneous agents. 
\par
In principle, it should be understandable how the macroeconomic dynamics results from the microscopic decisions and interactions on the level of producers and consumers \cite{Wolfgang,Aoki} (as it was possible in the past to derive micro-macro links for other systems with a complex dynamical behavior such as interactive vehicle traffic \cite{trafficanalytics}). It should also be comprehensible how the macroscopic level (the aggregate econonomic situation) feeds back on the microscopic level (the behavior of consumers and producers), and to understand the economy as a complex, adaptive, self-organizing system \cite{SantaFe,Blume}. Concepts from evolutionary theory \cite{EvolutionaryEconomics} and ecology \cite{May} appear to be particularly promising \cite{Evolu}. This, however, requires a recognition of the importance of heterogeneity for the system (see the the previous subsection). 
\par
The lack of ecological thinking implies not only that the sensitive network interdependencies between the various agents in an economic system (as well as minority solutions) are not properly valued. It also causes deficiencies in the development and implementation of a sustainable economic approach based on recycling and renewable resources. Today, forestry science is probably the best developed scientific discipline concerning sustainability concepts \cite{forestry}. Economic growth to maintain social welfare is a serious misconception. From other scientific disciplines, it is well known that stable pattern formation is also possible for a constant (and potentially sustainable) inflow of energy \cite{pattform,pattform2}.

\subsection{Optimization of system performance}\label{optimization} 

One of the great achievements of economics is that it has developed a multitude of methods to use scarce resources efficiently.  A conventional approach to this is optimization. In principle, there is nothing wrong about this approach. Nevertheless, there are a number of problems with the way it is usually applied.
\begin{enumerate}
\item One can only optimize for one goal at a time, while usually, one needs to meet several objectives. This is mostly addressed by weighting the different goals (objectives), by executing a hierarchy of optimization steps (through ranking and prioritization), or by applying a satisficing strategy (requiring a minimum performance for each goal) \cite{multigoaloptim,evoloptim}. However, when different optimization goals are in conflict with each other (such as maximizing the throughput and minimizing the queue length in a production system), a sophisticated time-dependent strategy may be needed \cite{selfcontrol}.
\item There is no unique rule what optimization goal should be chosen. Low costs? High profit? Best customer satisfaction? Large throughput? Competitive advantage? Resilience? \cite{BioLogistics} In fact, the choice of the optimization function is arbitrary to a certain extent and, therefore, the result of optimization may vary largely.  Goal selection requires strategic decisions, which may involve normative or moral factors (as in politics). In fact, one can often observe that, in the course of time, different goal functions are chosen. Moreover,
note that the maximization of certain objectives such as resilience or ``fitness'' depends not only on factors that are under the control of a company. Resilience and ``fitness'' are functions of the whole system, in particularly, they also depend on the competitors and the strategies chosen by them. 
\item The best solution may be the combination of two bad solutions and may, therefore, be overlooked. In other words,
there are ``evolutionary dead ends'', so that gradual optimization may not work. (This problem can be partially overcome by the application of evolutionary mechanisms \cite{evoloptim}). 
\item In certain systems (such as many transport, logistic, or production systems), optimization tends to drive the system towards instability, since the point of maximum efficiency is often in the neighborhood or even identical with the point of breakdown of performance. Such breakdowns in capacity or performance can result from inefficiencies due to dynamic interaction effects.  For example, when traffic flow reaches its maximum capacity, sooner or later it breaks down. As a consequence, the road capacity tends to drop during the time period where it is most urgently needed, namely during the rush hour \cite{stopandgo,capacitydrop}.
\item Optimization often eliminates reduncancies in the system and, thereby, increases the vulnerability to perturbations, i.e. it decreases robustness and resilience.
\item Optimization tends to eliminate heterogeneity in the system  \cite{Why}, while heterogeneity frequently supports adaptability and resilience.
\item Optimization is often performed with centralized concepts (e.g. by using supercomputers that process information collected all over the system). Such centralized systems are vulnerable to disturbances or failures of the central control unit. They are also sensitive to information overload, wrong selection of control parameters, and delays in adaptive feedback control. In contrast, decentralized control (with a certain degree of autonomy of local control units) may perform better, when the system is complex and composed of many heterogeneous elements, when the optimization problem is NP hard, the degree of fluctuations is large, and predictability is restricted to short time periods \cite{Windt,irgc}. Under such conditions, decentralized control strategies can perform well by adaptation to the actual local conditions, while being robust to perturbations. Urban traffic light control is a good example for this \cite{selfcontrol,Patent}.
\item Further on, today's concept of quality control appears to be awkward. It leads to a never-ending contest, requiring people and organizations to fulfil permanently increasing standards. This leads to over-emphasizing measured performance criteria, while non-measured success factors are neglected. The engagement into non-rewarded activities is discouraged, and innovation may be suppressed (e.g. when evaluating scientists by means of their h-index, which requires them to focus on a big research field that generates many citations in a short time). 

While so-called ``beauty contests'' are considered to produce the best results, they will eventually absorb more and more resources for this contest, while less and less time remains for the work that is actually to be performed, when the contest is won. Besides, a large number of competitors have to waste considerable resources for these contests which, of course, have to be paid by someone. In this way, private and public sectors (from physicians over hospitals, administrations, up to schools and universities) are aching under the evaluation-related administrative load, while little time remains to perform the work that the corresponding experts have been trained for. It seems na\"{\i}ve to believe that this would not waste resources. Rather than making use of individual strengths, which are highly heterogeneous, today's way of evaluating performance enforces a large degree of conformity. 
\end{enumerate}
There are also some problems with parameter fitting, a method based on
optimization as well. In this case, the goal function is typically an
error function or a likelihood function. Not only are calibration
methods often ``blindly'' applied in practice (by people who are not
experts in statistics), which can lead to overfitting (the fitting of
meaningless ``noise''), to the neglection of collinearities (implying
largely variable parameter values), or to inaccurate and problematic
parameter determinations (when the data set is insufficient in size,
for example, when large portfolios are to be optimized
\cite{Kondor}). As estimates for past data are not necessarily indicative for the future, 
making predictions with interpolation approaches can be quite problematic (see also Sec. \ref{equilibrium} for the challenge of time dependence). Moreover, classical calibration methods 
do not reveal inappropriate model specifications
(e.g. linear ones, when non-linear models would be needed, or unsuitable choices of
model variables). Finally, they do not identify unknown unknowns
(i.e. relevant explanatory variables, which have been overlooked in
the modeling process).

\subsection{Control approach}\label{control} 

Managing economic systems is a particular challenge, not only for the reasons discussed in the previous section. As large economic systems belong to the class of complex systems, they are hard or even impossible to manage with classical control approaches \cite{ManagComp,irgc}. 

Complex systems are characterized by a large number of system elements (e.g. individuals, companies, countries, ...), which have non-linear or network interactions causing mutual dependencies and responses. Such systems tend to behave dynamic rather than static and probabilistic rather than deterministic. They usually show a rich, hardly predictable, and sometimes paradoxical system behavior. Therefore, they challenge our way of thinking \cite{Doerner}, and their controllability is often overestimated (which is sometimes paraphrased as ``illusion of control'') \cite{Kempf,IllCont,Why}. In particular, causes and effects are typically not proportional to each other, which makes it difficult to predict the impact of a control attempt.

A complex system may be unresponsive to a control attempt, or the latter may lead to unexpected, large changes in the system behavior (so-called ``phase transitions'', ``regime shifts'', or ``catastrophes'') \cite{phasetrans}. The unresponsiveness is known as principle of Le Chatelier or Goodhart's law \cite{Good}, according to which a complex system tends to counteract external control attempts. However, regime shifts can occur, when the system gets close to so-called ``critical points'' (also known as ``tipping points''). Examples are sudden changes in public opinion (e.g. from pro to anti-war mood, from a smoking tolerance to a public smoking ban, or from buying energy-hungry sport utilities vehicles (SUVs) to buying environmentally-friendly cars). 
\par
Particularly in case of network interactions, big changes may have small, no, or unexpected effects. Feedback loops, unwanted side effects, and circuli vitiosi are quite typical. Delays may cause unstable system behavior (such as bull-whip effects) \cite{beergame}, and over-critical perturbations can create cascading failures \cite{Battiston}. Systemic breakdowns (such as large-scale blackouts, bankruptcy cascades, etc.) are often a result of such domino or avalanche effects \cite{irgc}, and their probability of occurrence as well as their resulting damage are usually underestimated. Further examples are epidemic spreading phenomena or disasters with an impact on the socio-economic system. A more detailed discussion is given in Refs. \cite{ManagComp,irgc}.
\par
Other factors contributing to the difficulty to manage economic systems are the large heterogeneity of system elements and the considerable level of randomness as well as the possibility of a chaotic or turbulent dynamics (see Sec. \ref{linear}). Furthermore, the agents in economic systems are responsive to information, which can create self-fulfilling or self-destroying prophecy effects. Inflation may be viewed as example of such an effect. Interestingly, in some cases one even does not know in advance, which of these effects will occur.
\par
It is also not obvious that the control mechanisms are well designed
from a cybernetic perspective, i.e. that we have sufficient
information about the system and suitable control variables to make
control feasible. For example, central banks do not have terribly many
options to influence the economic system. Among them are performing
open-market operations (to control money supply), adjustments in
fractional-reserve banking (keeping only a limited deposit, while
lending a large part of the assets to others), or adaptations in the
discount rate (the interest rate charged to banks for borrowing
short-term funds directly from a central bank). Nevertheless, the central banks are asked to meet multiple goals such as 
\begin{itemize}
\item to guarantee well-functioning and robust financial markets, 
\item to support economic growth, 
\item to balance between inflation and unemployment, 
\item to keep exchange rates within reasonable limits.
\end{itemize}
Furthermore, the one-dimensional variable of ``money'' is also used to influence {\it individual} behavior via taxes (by changing behavioral incentives). It is questionable, whether money can optimally meet all these goals at the same time (see Sec. \ref{optimization}). We believe that a computer, good food, friendship, social status, love, fairness, and knowledge can only to a certain extent be replaced by and traded against each other. Probably for this reason, social exchange comprises more  than just material exchange \cite{Fiske,seeLarsErik,Frey}. 
\par
It is conceivable that financial markets as well are trying to meet too many goals at the same time. This includes
\begin{itemize}
\item to match supply and demand,
\item to discover a fair price,
\item to raise the foreign direct investment (FDI),
\item to couple local economies with the international system,
\item to facilitate large-scale investments,
\item to boost development,
\item to share risk,
\item to support a robust economy, and
\item to create opportunities (to gamble, to become rich, etc.).
\end{itemize}
Therefore, it would be worth stuyding the system from a cybernetic control perspective. Maybe, it
would work better to separate some of these functions from each other rather than mixing them.

\subsection{Human factors}\label{human}

Another aspect that tends to be overlooked in mainstream economics is the relevance of psychological and social factors such as emotions, creativity, social norms, herding effects, etc. It would probably be wrong to interpret these effects just as a result
of perception biases (see Sec. \ref{economicus}). Most likely, these human factors serve certain functions such as supporting the creation of public goods \cite{Ostrom}  or collective intelligence \cite{collectiveintelligence,Swarm}. 

As Bruno Frey has pointed out, economics should be seen from a social science perspective \cite{Frey2}. In particular, research on happiness has revealed that there are more incentives than just financial ones that motivate people to work hard \cite{Frey}. Interestingly, there are quite a number of factors which promote volunteering \cite{seeLarsErik}. 
\par
It would also be misleading to judge emotions from the perspective of irrational behavior. They are a quite universal and a relatively energy-consuming way of signalling. Therefore, they are probably more reliable than non-emotional signals. Moreover, they create empathy and, consequently, stimulate mutual support and a readiness for compromises. It is quite likely that this creates a higher degree of cooperativeness in social dilemma situations and, thereby, a higher payoff on average as compared to emotionless decisions, which often have drawbacks later on.

\subsection{Information}\label{inform}

Finally, there is no good theory that would allow one to assess the relevance of information in economic systems. Most economic models do not consider information as an explanatory variable, although information is actually a stronger driving force of urban growth and social dynamics than energy \cite{Bettencourt}. While we have an information theory to determine the number of bits required to encode a message, we are lacking a theory, which would allow us to assess what kind of information is relevant or important, or what kind of information will change the social or economic world, or history. This may actually be largely dependent on the perception of pieces of information, and on normative or moral issues filtering or weighting information. Moreover, we lack theories describing what will happen in cases of coincidence or contradiction of several pieces of information. When pieces of information interact, this can change their interpretation and, thereby, the decisions and behaviors resulting from them. That is one of the reasons why socio-economic systems are so hard to predict: ``Unknown unknowns'', structural instabilities, and innovations cause emergent results and create a dynamics of surprise \cite{McDaniel}.

\section{Role of other scientific fields}

\subsection{Econophysics, ecology, computer science}\label{econophysics}

The problems discussed in the previous two sections pose interesting practical and fundamental challenges for economists, but also other disciplines interested in understanding economic systems. Econophysics, for example, pursues a physical approach to economic systems, applying methods from statistical physics \cite{Wolfgang}, network theory \cite{Barabasi,Schweitz}, and the theory of complex systems \cite{Weidlich2,Frank}. A contribution of physics appears quite natural, in fact, not only because of its tradition in detecting and modeling regularities in large data sets \cite{bigdata}. Physics also has a lot of experience how to theoretically deal with problems such as time-dependence, fluctuations, friction, entropy, non-linearity, strong interactions, correlations, heterogeneity, and many-particle simulations (which can be easily extended towards multi-agent simulations). In fact, physics has influenced economic modeling already in the past. Macroeconomic models, for example, were inspired by thermodynamics. More recent examples of relevant contributions by physicists concern models of self-organizing conventions \cite{convention}, of geographic agglomeration \cite{agglomeration}, of innovation spreading \cite{ebeling}, or of financial markets \cite{Econophysics}, to mention just a few examples. 
\par
One can probably say that physicists have been among the pioneers calling for new approaches in economics \cite{Wolfgang,Weidlich2,Econophysics,Minority,Bouchaud,Farmer,Buchanan}. A particularly visionary book beside Wolfgang Weidlich's work was the ``Introduction to Quantitative Aspects of Social Phenomena'' by Elliott W. Montroll and Wade W. Badger, which addressed by mathematical {\it and} empirical analysis subjects as diverse as population dynamics, the arms race, speculation patterns in stock markets, congestion in vehicular traffic as well as the problems of atmospheric pollution, city growth and developing countries already in 1974 \cite{Elliott}.
\par
Unfortunately, it is impossible in our paper to reflect the numerous contributions of the field of econophysics in any adequate way. The richness of scientific contributions is probably reflected best by the Econophysics Forum run by Yi-Cheng Zhang \cite{Fribourg}. Many econophysics solutions are interesting, but so far they are not broad and mighty enough to replace the rational agent paradigm with its large body of implications and applications. Nevertheless, considering the relatively small number of econophysicists, there have been many promising results. The probably largest fraction of publications in econophysics in the last years had a data-driven or computer modeling approach to financial markets \cite{Econophysics}. But econophyics has more to offer than the analysis of financial data (such as fluctuations in stock and foreign currency exchange markets), the creation of interaction models for stock markets, or the development of risk management strategies. Other scientists have focused on statistical laws underlying income and wealth distributions, non-linear market dynamics, macroeconomic production functions and conditions for economic growth or agglomeration, sustainable economic systems, business cycles, microeconomic interaction models, network models, the growth of companies, supply and production systems, logistic and transport networks, or innovation dynamics and diffusion. An overview of subjects is given, for example, by Ref. \cite{SocioEconophysics} and the contributions to annual spring workshop of the Physics of Socio-Economic Systems Division of the DPG \cite{FVSOE}.
\par
To the dissatisfaction of many econophysicists, the transfer of knowledge often did not work very well or, if so, has not been well recognized \cite{Roehner}. Besides scepticism on the side of many economists with regard to novel approaches introduced by ``outsiders'', the limited resonance and level of interdisciplinary exchange in the past was also caused in part by econophysicists. In many cases, questions have been answered, which no economist asked, rather than addressing puzzles economists are interested in. Apart from this, the econophysics work was not always presented in a way that linked it to the traditions of economics and pointed out deficiencies of existing models, highlighting the relevance of the new approach well. Typical responses are: Why has this model been proposed and not another one? Why has this simplification been used (e.g. an Ising model of interacting spins rather than a rational agent model)? Why are existing models not good enough to describe the same facts? What is the relevance of the work compared to previous publications? What practical implications does the finding have? What kind of paradigm shift does the approach imply? Can existing models be modified or extended in a way that solves the problem without requiring a paradigm shift? Correspondingly, there have been criticisms not only by mainstream economists, but also by colleagues, who are open to new approaches \cite{Worry}.
\par
Therefore, we would like to suggest to study the various economic subjects from the perspective of the above-mentioned fundamental challenges, and to contrast econophysics models with traditional economic models, showing that the latter leave out important features. It is important to demonstrate what properties of economic systems cannot be understood for fundamental reasons within the mainstream framework (i.e. cannot be dealt with by additional terms within the modeling class that is conventionally used). In other words, one needs to show why a paradigm shift is unavoidable, and this requires careful argumentation. We are not claiming that this has not been done in the past, but it certainly takes an additional effort to explain the essence of the econophysics approach in the language of economics, particularly as mainstream economics may not always provide suitable terms and frameworks to do this. This is particularly important, as the number of econophysicists is small compared to the number of economists, i.e. a minority wants to convince an established majority. To be taken seriously, one must also demonstrate a solid knowledge of related previous work of economists, to prevent the stereotypical reaction that the subject of the paper has been studied already long time ago (tacitly implying that it does not require another paper or model to address what has already been looked at before).
\par
A reasonable and promising strategy to address the above fundamental and practical challenges is to set up multi-disciplinary collaborations in order to combine the best of all relevant scientific methods and knowledge. It seems plausible that this will generate better models and higher impact than working in separation, and it will stimulate scientific innovation. Physicists can contribute with their experience in handling large data sets, in creating and simulating mathematical models, in developing useful approximations, in setting up laboratory experiments and measurement concepts. Current research activities in economics do not seem to put enough focus on
\begin{itemize}
\item modeling approaches for complex systems \cite{pluralistic},
\item computational modeling of what is not analytically tractable anymore, e.g. by agent-based models \cite{Leigh,Harrison,Davis},
\item testable predictions and their empirical or experimental validation \cite{Kagel},
\item managing complexity and systems engineering approaches to identify alternative ways of organizing financial markets and economic systems \cite{Lux,Salzano,ManagingComplexity}, and
\item an advance testing of the effectiveness, efficiency, safety, and systemic impact (side effects) of innovations, before they are implemented in economic systems. This is in sharp contrast to mechanical, electrical, nuclear, chemical and medical drug engineering, for example. 
\end{itemize}
Expanding the scope of economic thinking and paying more attention to these natural, computer and engineering science aspects will certainly help to address the theoretical and practical challenges posed by economic systems. Besides physics, we anticipate that also evolutionary biology, ecology, psychology, neuroscience, and artificial intelligence will be able to make significant contributions to the understanding of the roots of economic problems and how to solve them. In conclusion, there are interesting scientific times ahead.

\subsection{Social Sciences}\label{social}

It is a good question, whether answering the above list of fundamental challenges will sooner or later solve the practical problems as well. We think, this is a precondition, but it takes more, namely the consideration of social factors. In particular, the following questions need to be answered:
\begin{enumerate}
\item How to understand human decision-making? How to explain deviations from rational choice theory and the decision-theoretical paradoxes? Why are people risk averse?
\item How does consciousness and self-consciousness come about?
\item How to understand creativity and innovation?
\item How to explain homophily, i.e. the fact that individuals tend to agglomerate, interact with and imitate {\it similar} others?
\item How to explain social influence, collective decision making, opinion dynamics and voting behavior?
\item Why do individuals often cooperate in social dilemma situations?
\item How do indirect reciprocity, trust and reputation evolve? 
\item How do costly punishment, antisocial punishment, and discrimination come about?
\item How can the formation of social norms and conventions, social roles and socialization, conformity and integration be understood?
\item How do language and culture evolve? 
\item How to comprehend the formation of group identity and group dynamics? What are the laws of coalition formation, crowd behavior, and social movements? 
\item How to understand social networks, social structure, stratification, organizations and institutions?
\item How do social differentiation, specialization, inequality and segregation come about?
\item How to model deviance and crime, conflicts, violence, and wars?
\item How to understand social exchange, trading, and market dynamics?
\end{enumerate}
We think that, despite the large amount of research performed on these subjects, they are still not fully understood. The ultimate goal would be to formulate mathematical models, which would allow one to understand these issues as emergent phenomena based on first principles, e.g. as a result of (co-)evolutionary processes. Such first principles would be the basic facts of human capabilities and the kinds of interactions resulting from them, namely
\begin{enumerate}
\item birth, death, and reproduction,
\item the need of and competition for resources (such as food and water),
\item the ability to observe their environment (with different senses), 
\item the capability to memorize, learn, and imitate,
\item empathy and emotions,
\item signaling and communication abilities,
\item constructive (e.g. tool-making) and destructive (e.g. fighting) abilities,
\item mobility and (limited) carrying capacity,
\item the possibility of social and economic exchange.
\end{enumerate}
Such features can, in principle, be implemented in agent-based models \cite{Schelling,Hegselmann,Nigel,Macy,Sawyer,Ep}. Computer simulations
of many interacting agents would allow one to study the phenomena emerging in the resulting (artificial or) model societies, and to compare them with stylized facts \cite{Ep,QuaSo,Fortunato}. The main challenge, however, is not to program a seemingly realistic computer game. We are looking for scientific models, i.e. the underlying assumptions need to be validated, and this requires to link computer simulations with empirical and experimental research \cite{socialexperimenting}, and with massive (but privacy-respecting) mining of social interaction data \cite{bigdata}. In the ideal case, there would also be an analytical understanding in the end, as it has been recently gained for interactive driver behavior \cite{trafficanalytics}.

\subsection*{Acknowledgements} 

The authors are grateful for partial financial support by the ETH Competence Center ``Coping with Crises in Complex Socio-Economic Systems'' (CCSS) through ETH Research Grant CH1-01 08-2 and by the Future and Emerging Technologies programme FP7-COSI-ICT of the European Commission through the project Visioneer (grant no.: 248438). They would like to thank for feedbacks on the manuscript by Kenett Dror, Tobias Preis and Gabriele Tedeschi as well as for inspiring discussions during a Visioneer workshop in Zurich from January 13 to 15, 2010, involving, besides the authors, Stefano Battiston Guido Caldarelli, Anna Carbone, Giovanni Luca Ciampaglia, Andreas Flache, Imre Kondor, Sergi Lozano, Thomas Maillart, Amin Mazloumian, Tamara Mihaljev, Alexander Mikhailov, Ryan Murphy, Carlos Perez Roca, Stefan Reimann, Aki-Hiro Sato, Christian Schneider, Piotr Swistak, Gabriele Tedeschi, and Jiang Wu. Last but not least, we are grateful to Didier Sornette, Frank Schweitzer and Lars-Erik Cederman for providing some requested references.


\begin{thebibliography}{99}
\bibitem{note} For example, three-body planetary motion has deterministic chaotic solutions, although it is a problem in classical mechanics, where the equations of motion optimize a Lagrangian functional.
\end{thebibliography}
\end{document}